\documentclass[12pt]{article}

\usepackage[natbib=true,style=authoryear,sorting=none]{biblatex}
\usepackage[a4paper,top=2cm,bottom=2cm,left=2cm,right=2cm,marginparwidth=1.75cm]{geometry}

\usepackage{amsmath}
\usepackage{graphicx}
\usepackage[colorlinks=true, allcolors=blue]{hyperref}
\usepackage{authblk}

\title{\bf uGMRT observation of the unidentified PeVatron candidate LHAASO J2108+5157}
\author[1,2]{Gunindra Krishna Mahanta\thanks{guninmohantaba@gmail.com}}
\author[3]{Subhashis Roy}
\author[1]{Sagar Godambe}
\author[1,2]{Bitan Ghosal}
\author[1]{Nilay Bhatt}
\author[1,2]{Subir Bhattacharyya\thanks{subir.bhattacharyya@gmail.com}}

\affil[1]{\it Astrophysical Sciences Division, Bhabha Atomic Research Centre, Mumbai - 400085} 
\affil[2]{\it Homi Bhabha National Institute, Anushaktinagar, Mumbai - 400094} 
\affil[3]{\it National Centre for Radio Astrophysics, Pune University Campus, Post Bag 3, Ganeshkhind,Pune, 411007}

\begin{document}

\maketitle

\begin{abstract}
\noindent Recent observations by the Large High Altitude Air Shower Observatory (LHAASO) detected Ultra High Energy (UHE) photons in the range 100 TeV to 1.4 PeV from twelve sources including Crab nebula.  The detection of these photons demands the presence of at least PeV energy particle in the source. It is important to understand particle acceleration and radiation emission processes in such source. One of those twelve sources, LHAASO J2108+5157 does not show any association or counterparts at any other wavelength. In search of counterpart, we surveyed the region with Giant Metrewave Radio Telescope (GMRT) at 650 MHz frequency. GMRT observation revel radio emission from an extended source within the PSF of LHAASO which shows disk-jet morphology. Considering the spatial association and extent of the source, it is plausible that particle acceleration to PeV energies originates from this source.

\end{abstract}

\section{Introduction}

The search for the origin of Cosmic Rays(CRs) is one of the primary motivations for very high energy $\gamma$-ray astronomy. It is widely accepted that accelerators located inside our Galaxy must produce the bulk
of CRs detected at Earth\citep{Cristofari_2021}. It is generally argued that CRs with energies upto $\sim$ 1 PeV($10^{15}$ eV)is produced within our Galaxy. Therefore, the sources which can accelerate particles up to $\sim$PeV are of special interest as extreme galactic accelerators, called PeVatrons. The principal signature of both electron and proton PeVatrons is ultrahigh-energy (UHE) $\gamma$ radiation. In recent years HAWC observatory has reported the detection of $\gamma$-ray photons of energy $\geq$ 50 TeV from 9 galactic sources \citep{Abeysekara_2020,Albert_2020}. Several new candidates showing $\gamma$-ray emission above 100 TeV have been reported by Tibet AS$\gamma$ collaboration\citep{Amenomori_2019} and HAWC collaboration\citep{Abeysekara_2020}. But the era of PeVatron physics begins with the detection of Ultra High Energy (UHE) $\gamma$-ray photons from a dozen of galactic sources by the Large High Altitude Air Shower Observatory(LHAASO). These detections brought in new challenges in our understanding of the particle acceleration processes as well as the radiation emission mechanisms responsible for the production of photons in the TeV - PeV energy range. This in turn requires the identification of the nature of the sources and their behavior in other wavebands. Most of these UHE sources, so called ``PeVatron" have association with known VHE source like Pulsar Wind Nebula (PWN), Supernova remnants (SNR), microquasars etc \citep{LHAASO_cat1_2024}. But few sources does not have any counterpart. LHAASO J2108+5157 is one of such source, which does not have any counterpart in any other wavelength range.

\section{About 1LHAASO J2108+5157}
The source LHAASO J2108+5157 was first reported by \cite{12_Discovery_Cao2021} along with discovery of emission of UHE photons from 12 galactic source by the Large High Altitude Air Shower Observatory (LHAASO). This source is detected in the both energy range 25-100 TeV as well as $>$100 TeV by Kilometer  Square Array (KM2A) and Water Cherenkov Detector Array (WCDA) respectively with a significance of 9.6$\sigma$ and 8.5$\sigma$. Position of the source is reported as RA = $317^{\circ}.22 \pm 0^{\circ}.07_{stat}$, Dec = $+51^{\circ}.95 \pm 0^{\circ}.05_{stat}$.
In LHAASO 1st catalog\citep{LHAASO_cat1_2024} position of the source is updated and renamed the source as 1LHAASO~J2108+5153u which is 0$^\circ$.06 away from the reported position of the source LHAASO J2108+5157. \\

\noindent The source LHAASO J2108+5157 doesn't have any VHE counterpart as seen in VHE cat TeVcat. The source region is also observed by LST-1 and VERITAS, but no significant detection was claimed. While, recently HAWC have reported 7$\sigma$ detection of the source \citep{HAWC_VERITAS_PoS_2023}.
Fermi LAT analysis of the region revels the presence of a VHE source 4FGL J2108+5155e which falls within the LHAASO PSF at an angular separation of $0^{\circ}.13$ degree away from the source LHAASO J2108+5157 \citep{12_Discovery_Cao2021, Abe_2023}. Considering angular size and spectral behavior, physical association of the Fermi LAT source with the PeVatron candidate could not be established.\\

\noindent X-ray observation of the source region by Swift-XRT and XMM-Newton with an exposure of 4.7 ks and 13.6 ks yields upper limits\citep{Stroch_2013, Abe_2023}. Recent observation of CO spectral line by radio telescopesrevels the presence of various molecular clouds ([MML2017]2870, [FTK-MC]2022, [FTK-MC]J2108, [FTK-MC]HS) in the source region \citep{Miville_2017, FKT_MC_2023, FKT_MC_Paper1_2023}.

\noindent In the present work, we observed the PeVatron source LHAASO J2108+5157 at 650 MHz frequency with uGMRT to search for low frequency radio counterpart of the source. In this manuscript, we report and discuss the result of our observation.

\section{Result and discussion}

\begin{figure}
    \includegraphics[width=\textwidth]{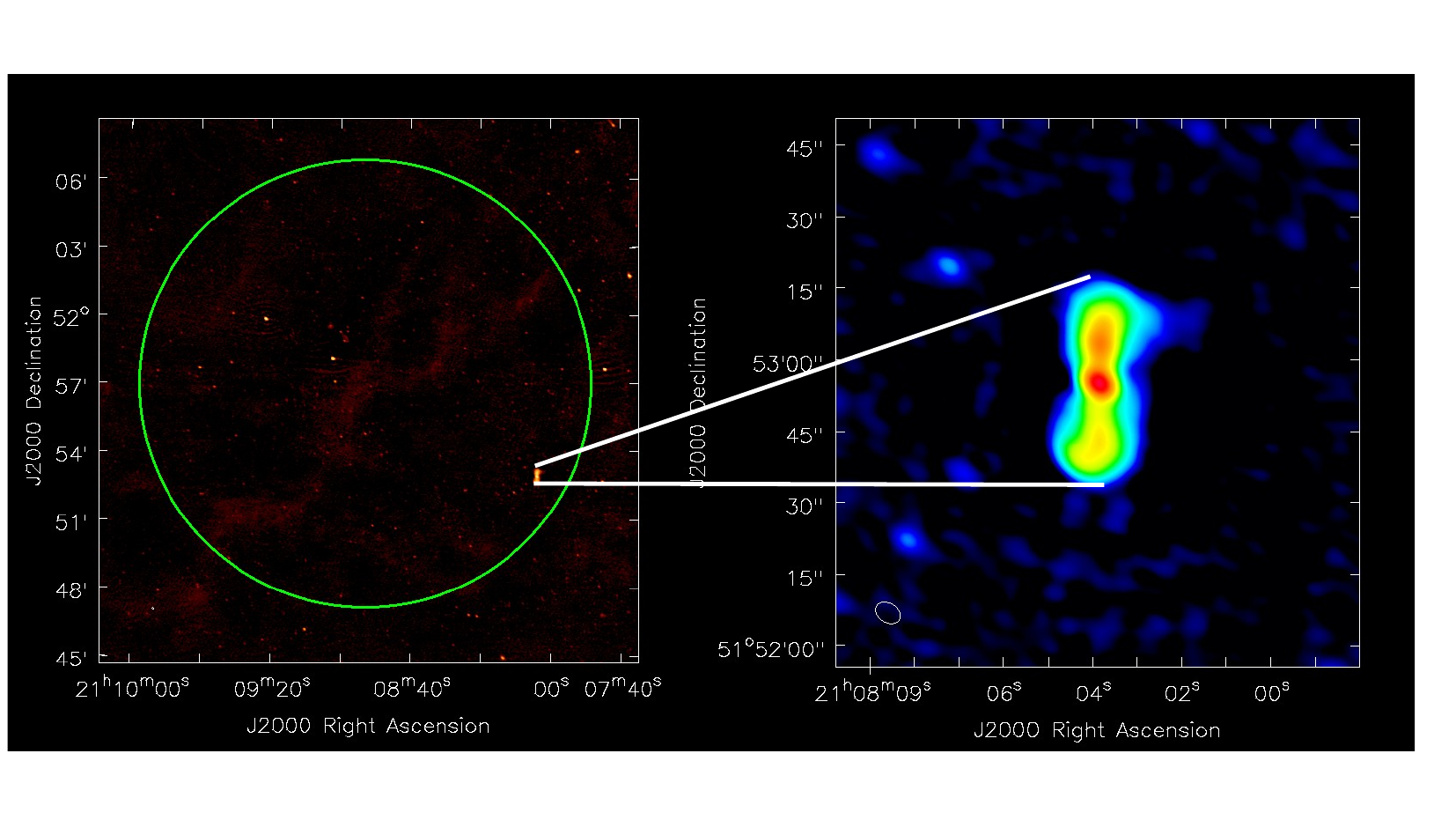}
    \caption{\textbf{Left:} Position of the source LHAASO J2108+5157 in the 650 MHz emission map of uGMRT. Position of the source as quoted by \cite{Discovery_J2108} is shown by green circle. \textbf{Right:} Morphology of the extended structure obtained in 650 MHz emission map. Synthesis beam is shown the white ellipse in the lower left corner of the emission map}
    \label{fig:Image2}
\end{figure}

\noindent
We observe the source LHAASO J2108+5157 through upgraded Giant Metrewave Radio telescope (uGMRT) on 2023 June 25 and 2023 August 11 with a total of 10 hours of exposure time. The observation was performed in continuum imaging mode covering 550-850 MHz frequency band (band 4).

\noindent The data were processed using standard technique\footnote{https://gmrt-tutorials.readthedocs.io/en/latest/index.html} with the Common Astronomy Software Applications (CASA) package of NRAO.

\noindent 
The result of 650 MHz uGMRT data observation is shown in Fig \ref{fig:Image2}. 
Green circle shows the upperlimit extension of the source LHAASO J2108+5157. 650 MHz emission map revels the presence of a lots of point source and one extended source (hereafter $\mathrm{GMRT_{ext}}$) within the region of interest (ROI). The SIMBAD\footnote{https://simbad.u-strasbg.fr/simbad/} and 2MASS\footnote{https://irsa.ipac.caltech.edu/Missions/2mass.html} database shows that most of these point source have spital association with stars, IR objects, and long period variables. But there is no clear indication about the nature of the extended source. NVSS, WENSS and CGPS catalog survey shows the presence of radio emission from the region of $\mathrm{GMRT_{ext}}$. But previous observations could not resolve the source structure due to limited angular resolution.

\noindent The source $\mathrm{GMRT_{ext}}$ also lies very close to the fermi source 4FGL J2108+5155e. The zoomed up image of $\mathrm{GMRT_{ext}}$ clearly shows a disk-jet like morphology in radio wave band. This types of morphology is very common in microquasar or radio galaxy. If it is a a radio galaxy, then it will be difficult to explain the observed UHE emission considering the CMB absorbtion. On the other hand microquasar is a possible PeVatron candidate \citep{microquasar_as_PeVatron1,
Mori_2023_Microquasar_as_PeVatron}. Presence of molecular cloud (MC) in this region hints towards a hadronic emission scenario, and protons from microquasar also can interact with the MCs to produce UHE gamma-rays. Hadronic nature of microquasar is also pointed out in various literature \citep{Vila_2010,Romero_2005,
Romero_2003}.

\section{Summary and conclusions}
\label{sec:conclusion}

\noindent 
We have reported the detection of an extended source in the region of dark PeVatron LHAASO J2108+5157. The extended source shows a disk-jet like morphology, which resembles to the radio image of a microquasar. If the source reported here is a microquasar, it can be a possible counterpart of the dark PeVatron LHAASO J2108+5157 reported by LHAASO. More observation of this source is required in different bands to completely understand the radiation emission mechanism of the source.

\section*{Acknowledgements}
\noindent 
\noindent 
We thank the staff of the GMRT who have made these observations possible. The GMRT is run by the National Centre for Radio
Astrophysics of the Tata Institute of Fundamental Research. 

\printbibliography

\end{document}